# NMR of liquid $^3$He in clay pores at 1.5 K

**Gazizulin R.R.[1], Klochkov A.V.[1], Kuzmin V.V.[1], Safiullin K.R.[1], Tagirov M.S.[1], Yudin A.N.[1], Izotov V.G.[2], Sitdikova L.M.[2]**

*[1] Physics Faculty, Kazan (Volga region) Federal University, 18 Kremlyovskaya St., Kazan 420008, Russia*

*[2] Geology Faculty, Kazan (Volga region) Federal University, 18 Kremlyovskaya St., Kazan 420008, Russia*

**Abstract**. In the present work a new method for studying porous media by nuclear magnetic resonance (NMR) of liquid $^3$He has been proposed. This method has been demonstrated in an example of a clay mineral sample. For the first time the integral porosity of clay sample has been measured. For investigated samples the value of integral porosity is in the range of 10-30 %. Inverse Laplace transform of $^3$He longitudinal magnetization recovery curve has been carried out, thus distribution of relaxation times $T_1$ has been obtained.

## 1 Introduction

Clay minerals are very important object for oil mining industry [1-2]. For a long time the clay minerals were assumed to be nonporous system, nontransparent for gases and liquids.

The study of clay minerals by traditional porometry methods are complicated by the unique properties of clay. The pore sizes are relatively small and probably closed for traditional porometry probes. Consequently, it is necessary to develop new methods for studying such materials.

The systems “liquid $^3$He – porous media” were previously studied intensively [3-7], but mainly for determination of $^3$He properties in contact with porous media and crystal powder samples.

It was found in [8] that the magnetic relaxation of the nuclear spins of liquid $^3$He in confined geometry acquires substantially new features as compared with relaxation in a bulk liquid and new mechanism of magnetic relaxation has been proposed by authors.

The properties of normal liquid $^3$He strongly depend on size of geometry where $^3$He is located. The main reason for that is highly effective spin diffusion, which allows seeing space restriction starting from several millimeter sizes. The nuclear magnetic relaxation of liquid $^3$He (both $T_1$ and $T_2$) takes place by means of fast spin diffusion from liquid $^3$He to adsorbed $^3$He and further effective surface relaxation in adsorbed layer of $^3He$ [3, 4, 6, 9]. Thus, the longitudinal relaxation time $T_1$, the transverse relaxation time $T_2$ and the spectral line width are strongly dependable on size of geometry filled by liquid $^3He$.

In the present work results of NMR measurements of liquid $^3He$ in pores of clay sample are reported and experimental data have been used to obtain the integral porosity of the sample.

## 2 The samples and methods

The clay sample was cut out of clay mineral in the shape of tablet with diameter of 5 mm and height $\Delta l$ = 3 mm. For calibration purposes a hole (diameter of 1.7 mm) has been drilled out in the center, along the tablet axis (axis ***l*** on Fig. 1), forming a calibration cavity. The ratio of the calibration cavity volume to the clay sample volume is 0.13. *A priori* NMR parameters of liquid $^3He$ in the calibration cavity will differ significantly from those in clay pores (Sect. 1).

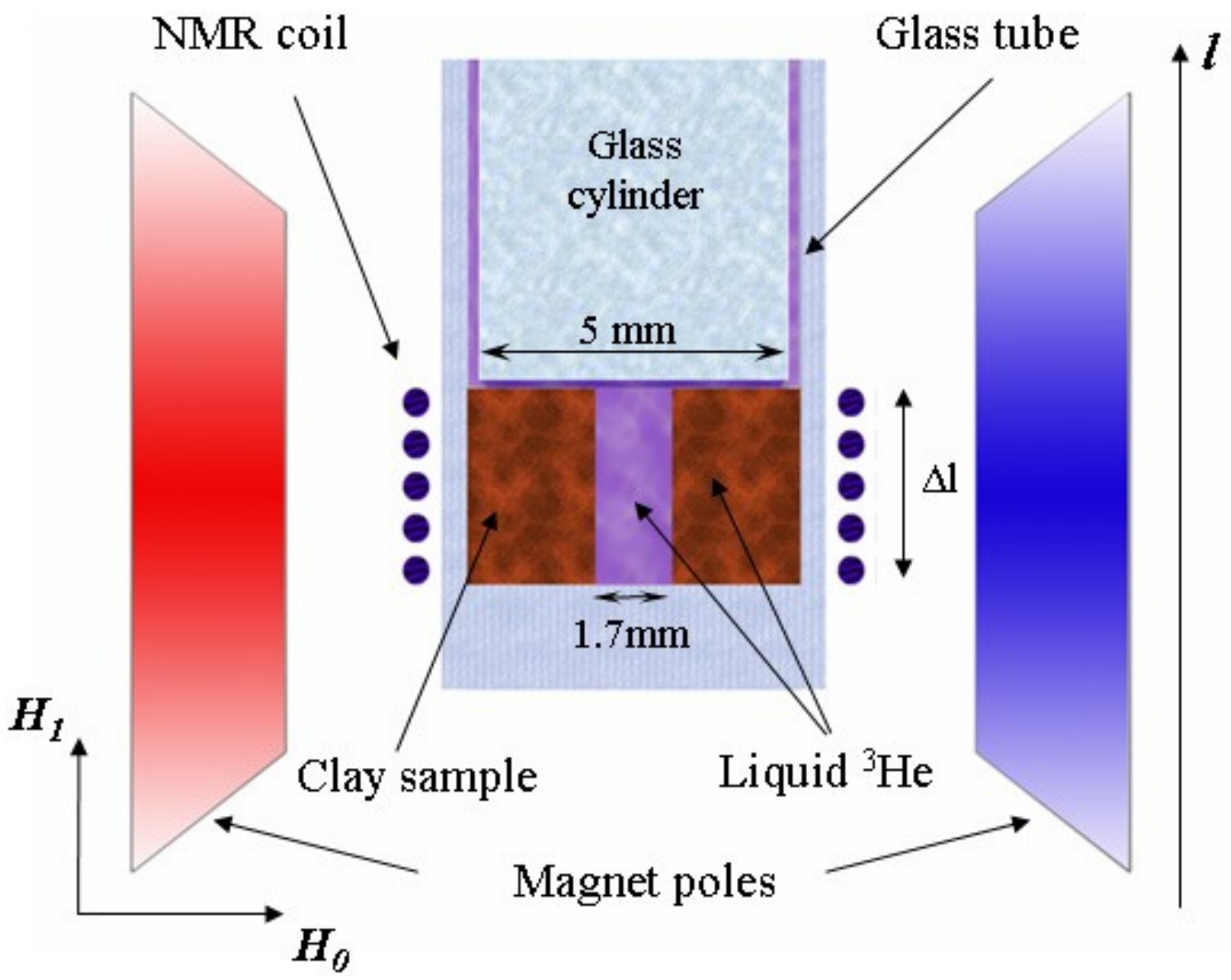

**Fig. 1.** Clay sample in a NMR probe.

The sample was placed in the glass tube which was sealed leak tight to the $^3He$ gas handling system. On the outer surface of the glass tube an NMR coil was mounted. The hand-made pulse NMR spectrometer has been used (frequency range 3 – 50 MHz). The pulse NMR spectrometer is equipped by resistive electrical magnet that has a magnetic field strength up to 1 T. The sample has been filled by $^3He$ under saturation vapor pressure at the temperature 1.5 K. Liquid $^3He$ inside of the calibration cavity was used as a reference for estimations of the integral porosity of the sample.

The longitudinal magnetization relaxation time $T_1$ of $^3$He was measured by the saturation recovery method using spin-echo signal. The observation of the spin-echo signal was possible due to specially induced inhomogeneity of the external magnetic field $H_0$. NMR probe has been decentered in the magnet by lifting along the axis ***l*** and, to a first approximation, a linear gradient of an external magnetic field appeared in the sample. The spin-spin relaxation time $T_2$ was measured by Hahn method.

## 3 Results and discussion

The system "clay sample – liquid $^3$He" was studied by measuring NMR spectra of liquid $^3$He as well as by studying nuclear spin kinetics of $^3$He on frequency 12 MHz and at temperature 1.5 K.

In Fig. 2 an example of liquid $^3$He NMR spectrum is presented.

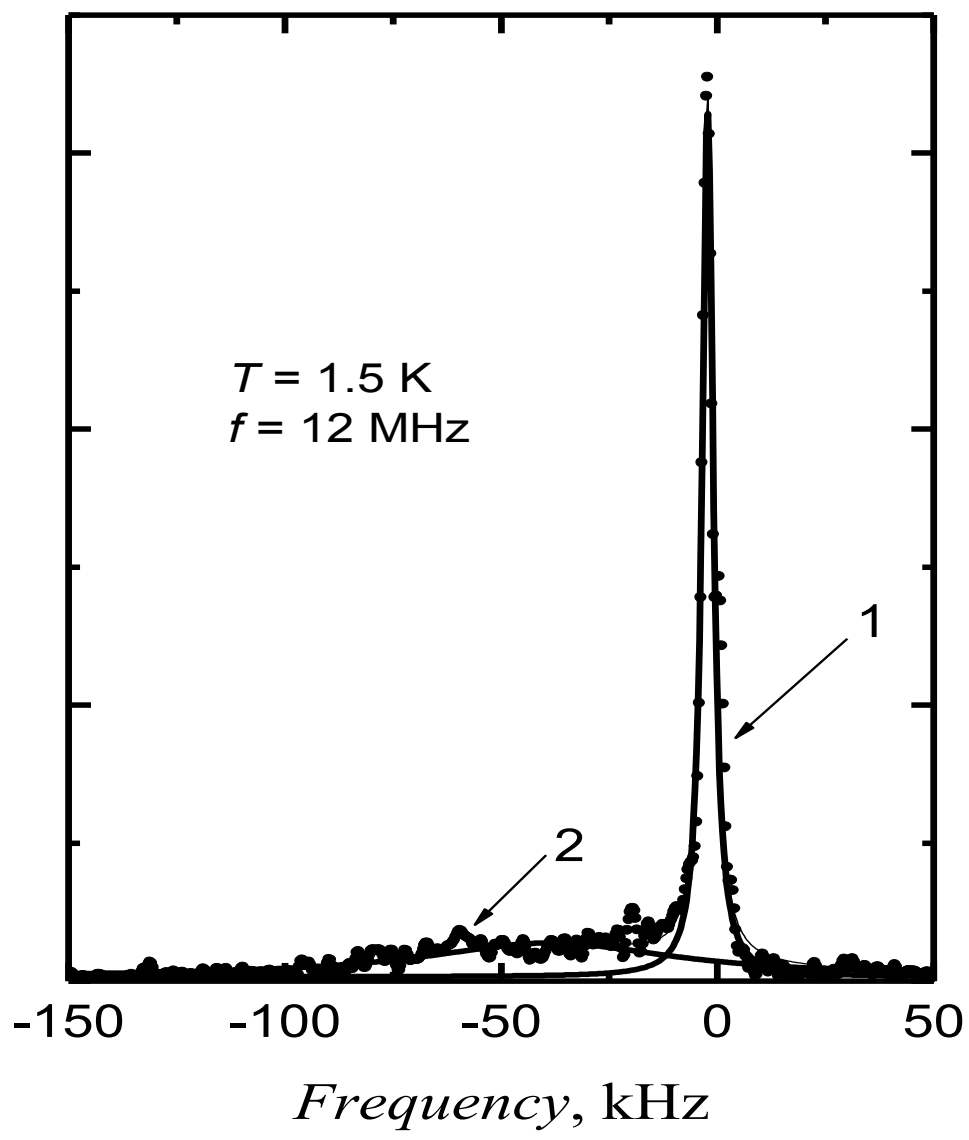

**Fig. 2.** The NMR spectrum of liquid $^3$He in the clay sample on the frequency 12 MHz at temperature 1.5 K (1 – liquid $^3$He signal from calibration cavity and 2 – liquid $^3$He signal from pores).

The NMR spectrum in the Fig. 2 was obtained by the Fast Fourier transformation of spin echo data from two channels of quadrature detector. Observed NMR spectrum can be described by two Lorentzian shape lines with widths of 3 and 91 kHz, and integral intensities 30200 ± 300 and 27500 ± 2800, correspondingly. As it was mentioned above in Sect. 1, the broad line *a priori* is the NMR signal of liquid $^3$He in pores of clay sample and the narrow line in the spectrum corresponds to $^3$He in a calibration cavity, because the average pore size in clay sample is much smaller than the size of the calibration cavity. The frequency shift of broad line indicates the presence of ferromagnetic impurities in the clay sample. It can be seen that the integral intensities of NMR signals of liquid $^3$He in calibration cavity and in clay pores are almost equal.

Nuclear spin relaxation times $T_1$ and $T_2$ of liquid $^3$He were measured on frequency 12 MHz and at the temperature 1.5 K. The behavior of the longitudinal magnetization recovery is non-exponential and consists of two processes (see example in Fig. 3).

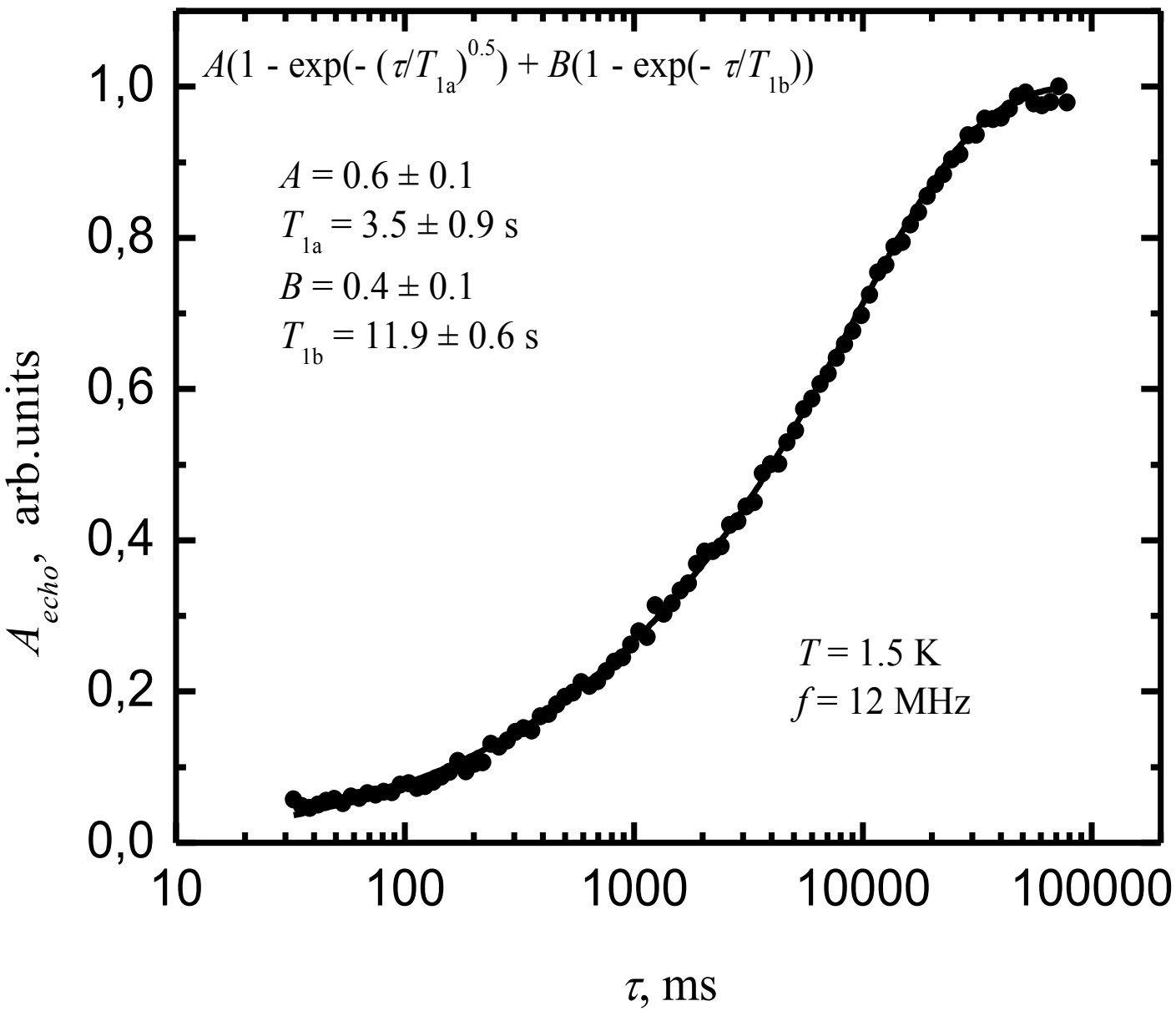

**Fig. 3.** Longitudinal magnetization recovery of liquid $^3$He nuclei in a clay sample on frequency 12 MHz and at temperature 1.5 K. Equation 1 describes experimental data with following parameters: $T_{1A} = 3.5$ s, $T_{1B} = 12$ s.

The longitudinal magnetization recovery curve can be described by following equation:

$$A_{echo} = A \cdot \left[ 1 - \exp\left( -\left( \frac{t}{T_{1a}} \right)^{0,5} \right) \right] + B \cdot \left[ 1 - \exp\left( -\frac{t}{T_{1b}} \right) \right], \qquad (1)$$

and parameters are: $T_{1a} = 3.5$ s, $T_{1b} = 12$ s, $A/B \approx 1$.

The first component in equation (1) with shorter $T_1$ corresponds to the nuclear magnetic relaxation of liquid $^3$He in pores of clay sample. There are several mechanisms of nuclear magnetic relaxation for liquids in porous media. In case of liquid $^3$He the main mechanism of nuclear magnetic relaxation is an energy transfer by means of anomalous fast spin diffusion from liquid $^3$He nuclei to adsorbed $^3$He nuclei and further effective surface relaxation in adsorbed layer of $^3He$ as a result of quantum exchange in 2-D film [6, 9]. Assuming single-exponential process of longitudinal relaxation in each pore, the power 0.5 in the first component of equation (1) indicates the presence of pore-size distribution.

The second component in (1) with longer $T_1$ corresponds to the nuclear magnetic relaxation of liquid $^3He$ in a calibration cavity (see Sect. 1). The observed value of $T_{1b}$ is much smaller than $T_1$ of bulk liquid $^3He$ (about 1,000 s [8]) because of fast molecular diffusion of $^3He$ molecules to the wall of the cavity. According to estimations, observed value $T_{1b}$ = 12 s corresponds to 0.6 mm radius of the diffusion, which comparable to the radius of the calibration cavity.

The behavior of the transverse magnetization decay is non-exponential, and also consists of two processes (Fig. 4).

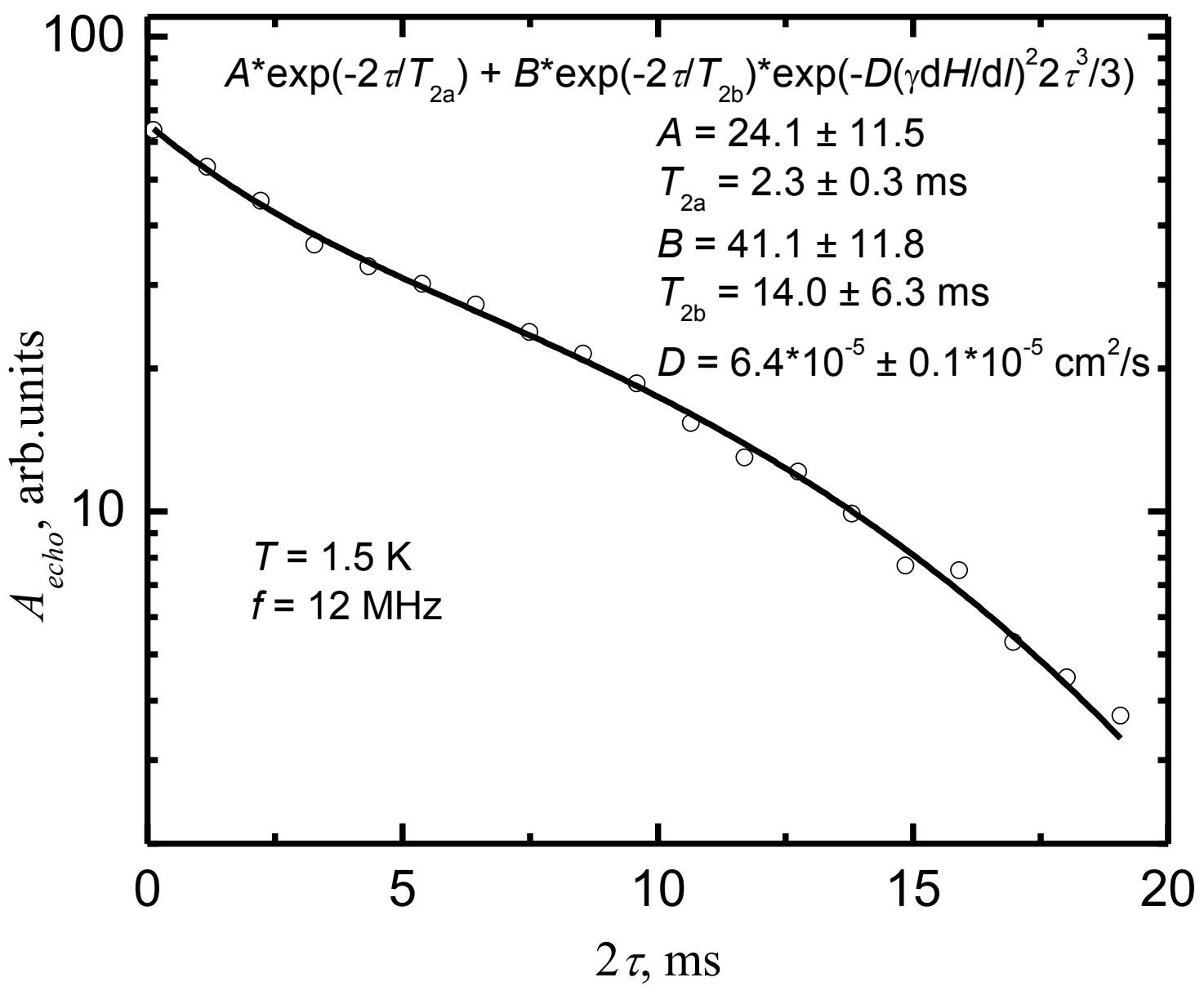

**Fig. 4.** Transverse magnetization decay of liquid $^3He$ nuclei in a clay sample on frequency 12 MHz and at the temperature 1.5 K. Equation 2 describes experimental data with following parameters: $T_{2A}$ = 2.3 ms, $T_{2B}$ = 14 ms, $D = 6.4\cdot10^{-5}$ cm$^2$/s.

The fast process of transverse magnetization decay is the relaxation of liquid $^3He$ in the pores of clay sample (see Sect. 1). The second process is the relaxation of liquid $^3He$ in the calibration cavity. One could expect equality of relaxation parameters $T_1$ and $T_2$ for liquid $^3He$ in a calibration cavity, as for bulk liquid. But, due to the presence of artificially induced external magnetic field gradient, phase relaxation through diffusion in an inhomogeneous magnetic field should be taken into account [10]. Consequently, the transverse magnetization decay curve is well described by:

$$A_{echo} = A \cdot \exp\left(-\frac{2t}{T_{2a}}\right) + B \cdot \exp\left(-\frac{2t}{T_{2b}}\right) \cdot \exp\left[-D\left(\gamma \frac{\partial H}{\partial l}\right)^2 \frac{2t^3}{3}\right], \qquad (2)$$

and parameters are: $T_{2a}$ = 2.3 ms, $T_{2b}$ = 14 ms, $D = 6.4 \cdot 10^{-5}$ cm$^2$/s, $A/B \approx 1$. For defining the value $\partial H/\partial l$ linear approach has been used. So $\partial H/\partial l \approx \Delta H/\Delta l$ where $\Delta H$ is the field inhomogeneity on the sample which was obtained from the width of the narrow line of the NMR spectrum (Fig. 2) and $\Delta l$ is the height of the sample (Fig. 1). The excellent agreement of diffusion coefficient $D$ value obtained by approximation of data (Fig. 4) with the experimental one from [11] proves applicability of proposed model for transverse relaxation of liquid $^3$He in an investigated system.

The inverse Laplace transform of longitudinal magnetization recovery curve for the detailed investigation of $^3$He relaxation in clay pores was performed using algorithm, based on a regularized inverse Laplace transform analysis with uniform penalty (UPEN) [12]. The multi-exponential inversion algorithm UPEN employs negative feedback to a regularization penalty to implement variable smoothing when both sharp and broad features appear on a single distribution of relaxation times. The inverse Laplace transform has been programmed using this algorithm. For the first time it has been applied to analysis of the magnetic relaxation of liquid $^3$He.

The distribution of relaxation times computed by UPEN is shown in Fig. 5, where narrow peak (in semi-log scale) corresponds to the signal of liquid $^3$He in the calibration cavity and wide peak with shorter $T_1$ values corresponds to liquid $^3$He in pores of clay sample.

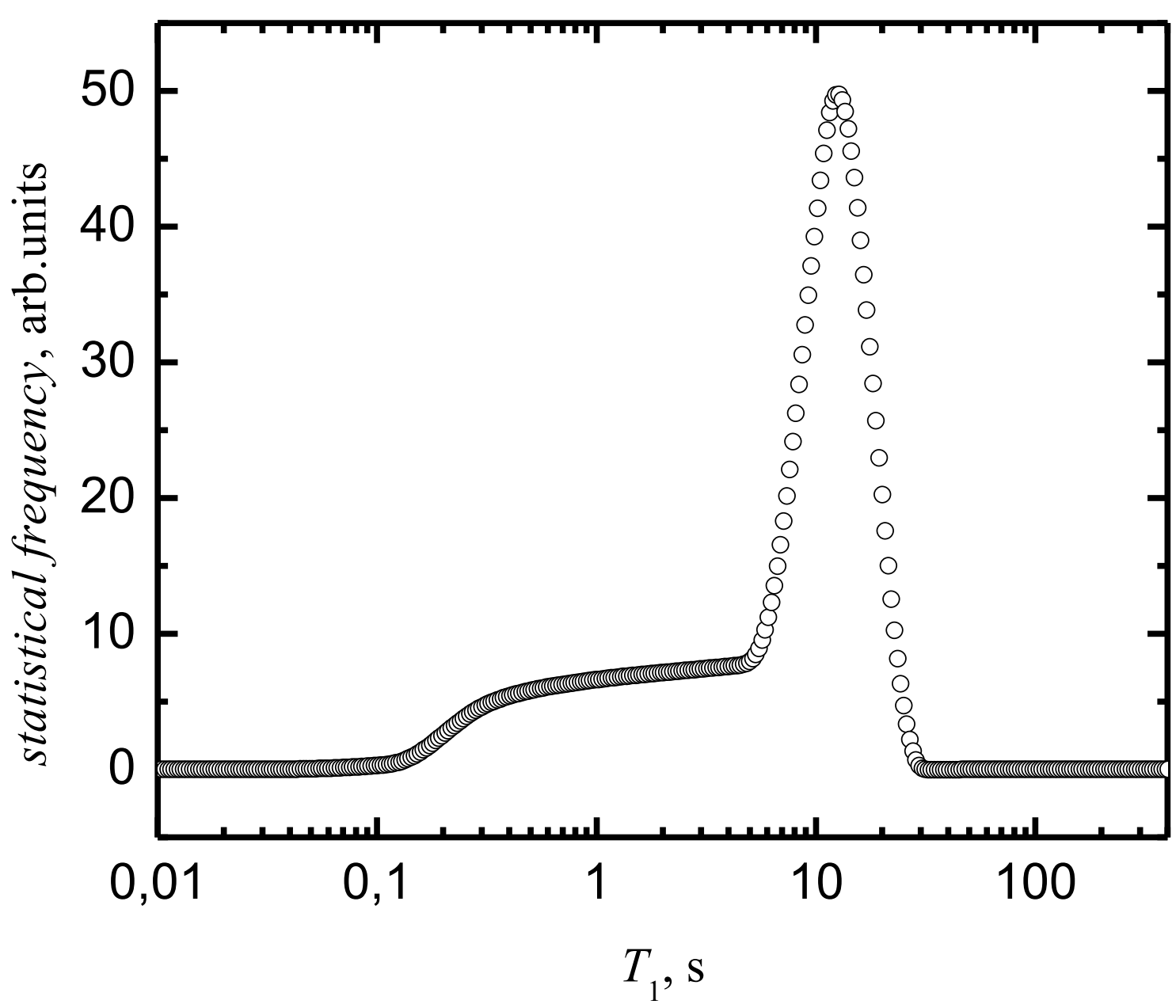

**Fig. 5.** Distribution of liquid $^3$He $T_1$ relaxation times for a clay sample obtained by regularized inverse Laplace transformation UPEN algorithm.

The distribution of relaxation times, obtained by inverse Laplace transform, could be converted into pore-size distribution, using specific models of relaxation. For instance, in case of very small pore sizes, when the $^3$He molecular diffusion length during the time of an experiment is longer than the size of pores, the longitudinal relaxation of liquid $^3$He would be single-exponential in each pore. The $T_1$ relaxation mechanism in this case would be surface relaxation, and direct knowledge of surface relaxation time $T_{1s}$, would allow obtaining pore size distribution. The surface relaxation itself could have different mechanisms, specially complicated by using geological sample, like in presented case.

The image of the clay sample obtained by scanning electron microscope Philips XL30 ESEM is presented in Fig. 6, where pores with different sizes can also be seen.

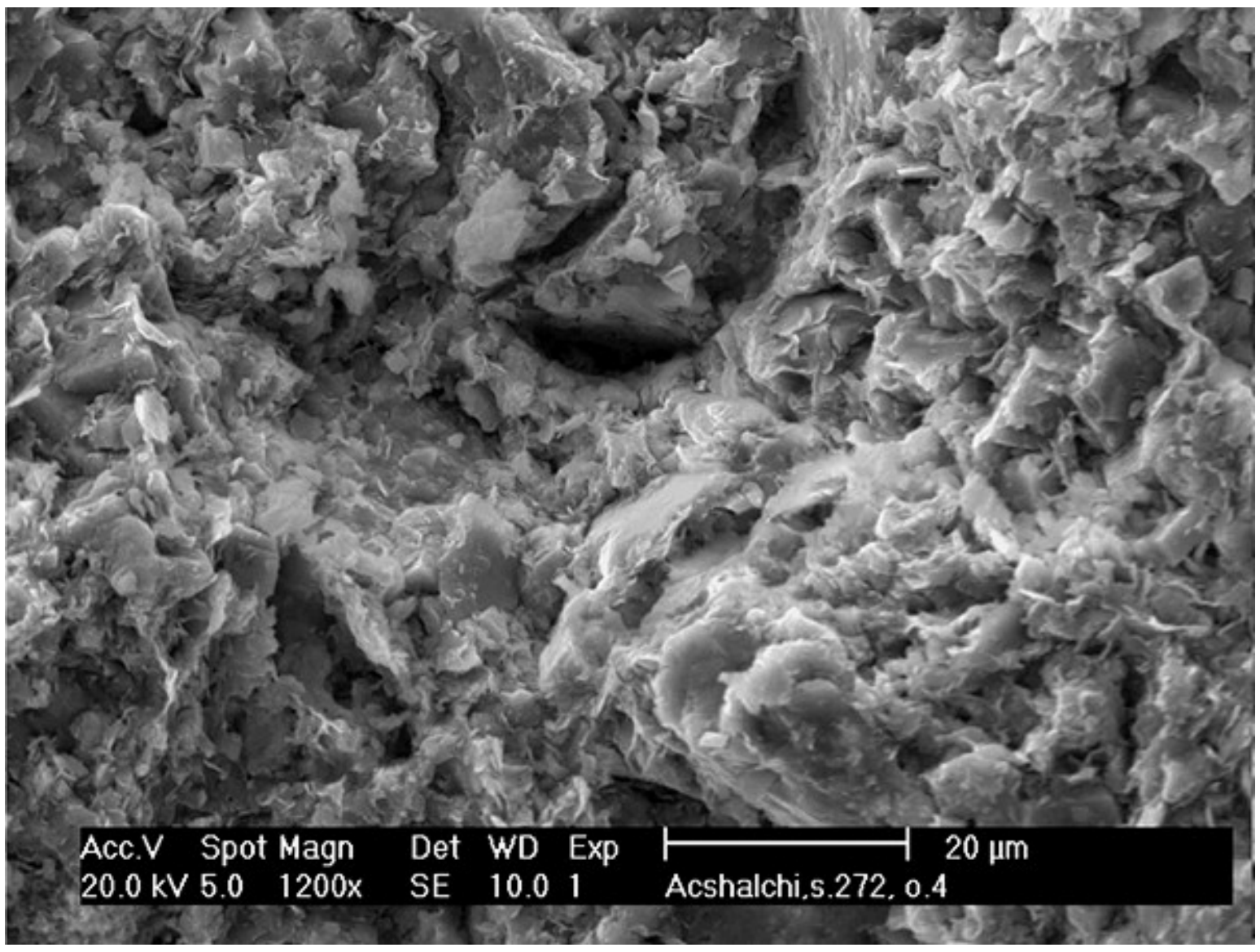

**Fig. 6.** Photo of clay sample by scanning electron microscope Philips XL30 ESEM.

In all obtained experimental data (NMR spectra, nuclear spin relaxation processes) the ratio between signals of liquid $^3$He in the calibration cavity and in pores of clay sample found to be close to 1. Taking into account the fact that the ratio of the calibration cavity volume to the clay sample volume is about 0.13, it can be found out that the sample porosity is about 13%.

## 4 Conclusion

The new method for study porous media by NMR of liquid $^3$He has been proposed in an example of clay sample. For the first time the integral porosity of clay sample has been measured. Combining all experimental results it can be concluded, that the porosity of a sample is 13%. The scanning electron microscopy shows the presence in the sample pores with different sizes.

It has been also demonstrated that inverse Laplace transform of $^3$He relaxation curves in porous media can give information about the distribution of relaxation times. The distribution of relaxation times could be converted into pore-size distribution, using specific models of relaxation.

**Acknowledgements**. The authors acknowledge A.V. Egorov for enlightening discussions.